\documentclass[preprint2]{aastex}

\begin{document}

\shorttitle{Nearby dwarf galaxy NGC~4214}
\shortauthors{I.Drozdovsky, R.Schulte-Ladbeck et al.}


\title{The Dwarf Irregular/Wolf-Rayet Galaxy NGC~4214: I. A New Distance,
Stellar Content, and Global Parameters\thanks{
Based on observations made with
the NASA/ESA {\em Hubble Space Telescope}, obtained at the Space Telescope
Science Institute, which is operated by the Association of Universities
for Research in Astronomy, Inc., Under NASA contract NAS 5-26555}}


\author{Igor O. Drozdovsky\thanks{
Astronomical Institute, St.Petersburg State
University, Petrodvoretz, 198904, Russia}\phantom{,}
 and Regina E. Schulte-Ladbeck}
\affil{University of Pittsburgh, Pittsburgh, PA 15260, USA}
\email{dio@phyast.pitt.edu, rsl@phyast.pitt.edu}

\author{Ulrich Hopp and Laura Greggio\thanks{Osservatorio Astronomico di Bologna, Bologna, Italy}}
\affil{Universit\"{a}tssternwarte M\"{u}nchen, M\"{u}nchen, FRG}
\email{hopp@usm.uni-muenchen.de, greggio@usm.uni-muenchen.de}

\and
\author{Mary M. Crone}
\affil{Skidmore College, Saratoga Springs, NY 12866, USA}
\email{mcrone@skidmore.edu}

\begin{abstract}
We present the results of a detailed
optical and near-IR
study of the nearby star-forming dwarf galaxy NGC~4214.
We discuss the stellar content, drawing particular attention
to the intermediate-age and/or old field stars, which are used as a
distance indicator.
On images obtained with the {\em Hubble Space Telescope\/} WFPC2 and NICMOS 
instruments in the equivalents of the $V$, $R$, $I$, $J$ and $H$ bands, 
the galaxy is well resolved into stars. We achieve limiting
magnitudes of $F814W \approx 27$ in the WF chips and $F110W \approx 25$ 
in the NIC2.
The optical and near-infrared color-magnitude diagrams confirm a 
core-halo galaxy morphology: 
an inner high surface-brightness
young population
within $\sim1\farcm5$ ($\sim1$ kpc) from the center of the galaxy, where the stars are
concentrated in bright complexes along a bar-like structure;
and a relatively low-surface-brightness, field-star population
extending out to at least $8\arcmin$ (7 kpc). The color-magnitude
diagrams of the core region show evidence of blue
and red supergiants, main-sequence stars, asymptotic giant branch
stars and blue loop stars.
We identify some candidate carbon stars from their extreme
near-IR color. The field-star population is dominated by
the ``red tangle", which contains the red giant branch.
We use the $I$-band luminosity
function to determine the distance based
on the tip-of-the-red-giant-branch method:  
$2.7\pm0.3$~Mpc. This is much closer 
than the values usually assumed in the literature,
and we provide revised distance dependent parameters such
as physical size, luminosity, H{\sc i} mass and star-formation rate.
From the mean color of the red giant branch in $V$ and $I$,
we estimate the mean metal abundance of this population as
[Fe/H]~$\simeq-1.7$~dex, with a large internal abundance spread 
characterized by $\sigma_{\rm int}({\rm [Fe/H]}) \approx 1$~dex.

\end{abstract}



\section{Introduction.}

Recent advances in the study of starburst galaxies
at high redshift emphasize the importance of
investigating the starburst phenomenon in the local Universe.
Lyman-break galaxies (LBGs), for instance, were discovered via a
pronounced Lyman break seen in the spectra of galaxies
whose far-UV spectra are dominated by emission from massive, young 
stars \citep{Steidel96a}. LBGs indeed show many spectroscopic similarities,
from
the far-UV throughout the optical range, with star-forming galaxies locally
\citep[cf. also,][]{Pettini01}. Steidel et al. (1996~a,b) compared rest-frame
UV spectra of LBGs with the UV spectrum of the central star-forming region
(NGC~4214-{\sc i}) in the local Irregular/Wolf-Rayet galaxy NGC~4214
\citep{Leitherer96}, and point out the striking likeness.

We here present a detailed study of the stellar populations in NGC~4214.
This galaxy is spatially so well resolved with the HST that individual
stars can be directly observed and placed on a color-magnitude diagram (CMD),
the observer's analog of the Hertzsprung-Russell diagram.
This paper is devoted to the characterization of the stellar
content of NGC~4214, and to the derivation of its
distance via the tip-of-the-red-giant-branch (TRGB)
method \citep{lee}. We also make some general inferences
about the star-formation history (SFH) of this galaxy. 

\begin{figure*}
\centering{
\vbox{\includegraphics{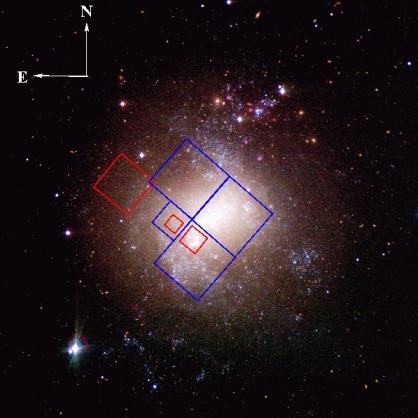}}\par
\vspace*{16.0cm}
}
\caption{Color $9\arcmin\times9\arcmin$ image of NGC~4214
from an RGB-combination of public $B$, $V$, $R$-band images obtained with
the Isaac Newton Telescope. The field of view  of the WFPC2 is shown in blue,
and that of NICMOS is illustrated in red.
\label{f:int_ima}
}
\end{figure*}

NGC~4214 is a high optical surface brightness Magellanic Irregular galaxy
(see NED) toward the Canes Venatici cloud
\citep{deVaucouleurs75}.
It is moderately metal-deficient, with a nebular abundance 
$12+\log({\rm O}/{\rm H})=8.20\div8.34$ \citep{Kobulnicky96},
or about $Z_{\sun}/4$ using $12+\log({\rm O}/{\rm H})=8.83$ from
\citet{Grevesse} for the Sun's Oxygen abundance. 
Morphologically, it consists of multiple blue star formation regions along a
central bar-like structure
surrounded by a large disk of gas which is seen in radio
H{\sc i}-observations as having a diameter of up to $\sim 15\arcmin$
\citep{Allsopp79}.
The star-formation regions give the impression of a
multi-arm spiral structure (see Fig.~\ref{f:int_ima}), but whether 
NGC~4214 is a
dwarf Spiral
remains unclear considering its complex velocity
field \citep{McIntyre97}. NGC~4214 is relatively gas rich even for a
Magellanic Irregular galaxy \citep{Sargent91}, and it has unusually blue
colors ($(B-V)_0=0.46$, NED).

\begin{figure*}
\vbox{\includegraphics{dio.fig2a.eps}
     {\includegraphics{dio.fig2a.eps}}\par
\vspace*{8.0cm}
}
\caption{NIC2 images of NGC~4214-{\sc ii}.
{\em Left} is the $F110W$ band image, {\em right} is the $F160W$ band image.
\label{f:nic2_ima}
}
\end{figure*}

\begin{figure*}
\centering{
\vbox{\includegraphics{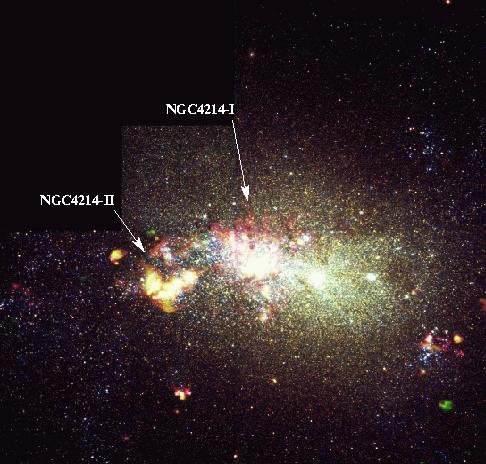}}\par
\vspace*{18.0cm}
}
\caption{WFPC2 image of NGC~4214 in the
$U+V+R+I+H\alpha+[O${\sc iii}] bands.
\label{f:wfpc2_ima}
}
\end{figure*}

NGC~4214 has been studied extensively in the last few years.
Two main star-formation regions have been identified in the central parts
of the galaxy: the larger NGC~4214-{\sc i} and the smaller, slightly younger
NGC~4214-{\sc ii}
(Fig.~\ref{f:wfpc2_ima}; Ma\'{\i}z-Apell\'aniz et al. \citeyear{Maiz99}
and references therein).
NGC~4214-{\sc i} is also known as the NW complex, and
NGC~4214-{\sc ii}, as the SE complex. These complexes
are rich in Wolf-Rayet stars of both the WN and WC sub-types, which trace
very recent star formation \citep{Sargent91,Mas99}. 
Spectral synthesis models suggest that the nuclear starburst,
NGC~4214-{\sc i}, is about
4~Myr old while NGC~4214-{\sc ii} is about $0.5-1$~Myr younger
\citep{Leitherer96,Mas99}.
\citet{Thronson88} used near- and far-infrared continuum and
CO emission line observations to study the star-formation properties of the
galaxy, and suggested that its present rate,
$0.5-1 \,{\rm M}_{\sun}{\rm yr}^{-1}$, is somewhat
higher than that averaged over a Hubble time. 
These conclusions are in agreement with FUV Ultraviolet Imaging Telescope
observations by \citet{Fanelli97}.

Past distance estimates for NGC~4214 based on its recession velocity
have varied from 3.6 to 7~Mpc
\citep{Allsopp79,Thronson88,Leitherer96,Martin98}. 
\citet{lsaz97} resolved the brightest blue supergiants (BSG) in a
small outlying stellar complex of NGC~4214 with ground-based optical imaging,
and derived a distance of 4.1~Mpc. It is well known that such distance
estimates suffer from large uncertainties. The new shorter distance
derived in this paper, on the other hand, is based on the numerous and well
resolved red-giant-branch (RGB) stars.
\citet{hopp99} first used near-IR photometry of RGB
stars with HST/NICMOS to revise the distance of NGC~4124 downward, 
to about 2~Mpc. An uncertainty in their distance determination
came from the lack of a calibration for the infrared tip of the RGB (TRGB) 
as a distance indicator. However, a small distance to NGC 4214,
$\approx$2.7~Mpc, also resulted when \citet{dio01}
employed the well-calibrated I-band TRGB method based on HST/WFPC2 data.
\citet{Maiz02} confirmed the results from the I-band TRGB method
when they determined 
a distance of $2.94 \pm 0.18$~Mpc based on a similar set of HST/WFPC2 data.
In this paper, we combine optical and near-IR HST data to investigate
the location of the TRGB in NGC~4214, to revisit the question of the
calibration of the near-IR TRGB as a distance indicator, and to propose
a distance of $2.7 \pm 0.3$  Mpc for NGC~4214.
We emphasize that the new distance, 2.7~Mpc,
results in a substantial revision in all distance-dependent
parameters of NGC~4214.

We present and discuss a deep optical-infrared study
of the stellar populations of NGC~4214. We use data from two {\sl HST}
instruments, NICMOS and WFPC2.
Multi-color stellar photometry provides us with the opportunity to 
study directly the 
resolved stellar populations of different masses, ages and chemical
abundances. The analysis of the stellar distribution on CMDs
is the most powerful way to investigate population
fractions and their spatial variations, to estimate the galaxy distance,
and to provide clues to its star formation history. 

\section{Observations and reductions.} \label{observations}

We combine single-star photometry in the optical and near-IR
to investigate the distance and stellar content of NGC~4214. 
Figure~\ref{f:int_ima} shows the location of the WFPC2
and NICMOS fields considered in this study 
superimposed on a color reproduction of an archival Isaac Newton
Telescope image. The Holmberg radius of NGC~4214 is
$10\farcm6\times10\farcm6$ \citep{Allsopp79}.  Therefore, 
the stellar halo probably spreads even farther than the
$9\arcmin\times9\arcmin$ field in Figure~\ref{f:int_ima}.

The $J$, $H$ NICMOS observations were gathered as part of a proposal by our
team aimed at studying any putative older stellar populations
in star-forming dwarf galaxies.
The advantage of using NICMOS to search for bright 
inter\-medi\-ate\--age/old stars
is that their spectral energy distributions peak in the near-IR.
Another advantage is the smaller extinction of near-IR compared to optical or
UV radiation.  However,
a drawback is that the field of the NIC2 camera, on which we concentrate
here, is rather small.
Figure~\ref{f:nic2_ima} presents $F110W$ and $F160W$ band images of
NGC~4214; these images cover the young star-forming region NGC~4214-{\sc ii}.

We also make extensive use of archival WFPC2 observations in $V$, $R$, $I$,
$H\alpha$, and $[O\,${\sc iii}$]$.
These data are precious because the $I$-band TRGB is
a well-calibrated distance indicator, and any interpretation of stellar
content depends critically on an accurate knowledge of the distance. 
We furthermore use the $I$-band TRGB to bootstrap a calibration for the
$H$-band TRGB as an extragalactic distance indicator. The WFPC2 camera has a
large field of view, and this allows us to compare spatial variations
of different stellar populations. 
Figure~\ref{f:wfpc2_ima} is a color
$(F814W+F702W+F656N)+(F555W+F502N)+(F336W)=(Red)+(Green)+(Blue)$
image of NGC~4214 obtained with the WFPC2, made in a similar way as the
Hubble Heritage team's image of this
galaxy\footnote{\anchor{http://www.stsci.edu/~jmaiz/ngc4214E.html}
{http://www.stsci.edu/$^\sim$jmaiz/ngc4214E.html}}.

\subsection{NICMOS}

We obtained NICMOS observations of NGC~4214 on 1998 July 23
as a part of GO program 7859\footnote{
Information about the observations can be gleaned directly from the
STScI WWW pages linked to the program ID.}.
The NICMOS instrument houses three cameras: NIC1, NIC2, and NIC3, in a
linear arrangement. Due to its low sensitivity and very small field of view,
the NIC1 camera produced images which contain few point sources.
The NIC3 images exhibit several
bright stars, but the stellar images are out of focus. Therefore, we 
discuss in this paper only the NIC2 data.

\begin{figure}[hbt]
\vbox{\includegraphics{dio.fig4a.eps}}
\vspace{2.9cm}
\vbox{\includegraphics{dio.fig4b.eps}}
\vspace{4.2cm}
\caption{{\tt ALLSTAR} residuals of the fitting of the
PSF models are shown as a function of near-IR magnitude.
Only stars included in the final photometric
list of NGC~4214/NIC2 are plotted.
\label{f:ST}
}
\end{figure}

\begin{figure}[hbt]
\vbox{\includegraphics{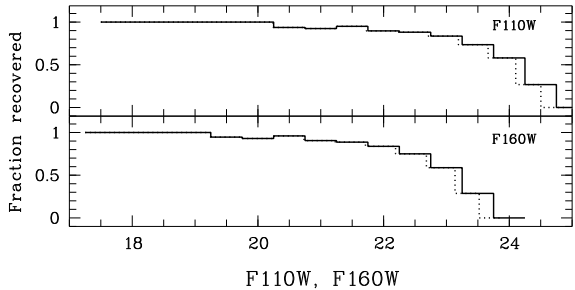}}
\vspace{4.3cm}
\caption{Completeness levels of the NIC2 photometry based on
artificial star tests with colors
$[0\fm5\leqslant(F110W-F160W)\leqslant1\fm5]$.
The solid line is the completeness level as a function of the input magnitude
with 0.5 mag bins.
The dashed line is the histogram corrected to be a function of the detected
magnitude.
Because of Malmquist bias
we detected some stars with
an input magnitude fainter than the detection limits; the noise scatters
these stars into brighter magnitudes.
\label{f:cmpl_nic2}
}
\end{figure}

The NIC2 camera, which has a field
of view of $19\farcs2 \times 19\farcs2$, was centered at
(J2000) R.A. $12^{\rm h}15^{\rm m}41\fs1$ and 
Dec. $+36\degr19\arcmin05\farcs8$ on NGC~4214-{\sc ii},
the smaller and younger of the two prominent star-forming regions.
Total integration times were about 5376~s in $F110W$, and 2688~s
in $F160W$.

We re-ran the STSDAS pipeline calibration with the latest
improved calibration files and software available in the standard NICMOS
{\tt CALNICA} pipeline. This software performs bias subtraction, dark-count
correction, flat-fielding, masking of bad pixels, and correction of
defective columns (bars).
Our NIC2 images are almost unaffected by what is known as the
``pedestal'' problem (an offset between the four quadrants of the camera),
so we did not need to correct for this effect.
We combined individual dithered images into a mosaic with the
{\tt DitherII} package, which performs a ``drizzling'' (a variable-pixel
linear reconstruction), correction for geometric distortion and removal
of cosmic rays \citep{Fruchter}.

Photometric processing was performed with the {\tt DAOPHOT} and
{\tt ALLSTAR} programs \citep{stetson} run within MIDAS.
We applied background-smoothing before conducting stellar
photometry. A master frame was produced using both the
$F110W$ and $F160W$ images. 
Two {\tt FIND/ALLSTAR/\linebreak[0]SUBSTAR} passes were used to identify
stars in this master frame. The resulting list of stellar coordinates was
given to ALLSTAR to perform the photometry in the individual 
$F110W$ and $F160W$ frames
using a PSF that quadratically varied with position in the each 
frame.

The NIC2 images exhibit diffraction rings around the stellar images
(see Fig.~\ref{f:nic2_ima}).
For the {\tt DAOPHOT} photometry we made a PSF radius large enough to
include all the visible diffraction rings, but used a fitting radius equal
to the radius of the first Airy minimum.

We merged the two $F110W$ and $F160W$ star lists requiring
a positional source coincidence of better than $0.5 \times FWHM$, or a two
pixels box size. Figure~\ref{f:ST} shows the error distributions for the
photometry.

We derived an aperture correction to the $0\farcs5$ standard aperture
from the data.
The final photometry uses the photometric
zero-points in the {\sl HST\/} Vegamag system as provided by the NICMOS team.
Subsequent CMDs in the ground-system ($J$, $H$) use the transformations which
we derived in SHGC99 from the NICMOS team's information about standard-star
observations on their photometry WWW site.

Completeness tests were performed using the usual procedure of artificial
star trials \citep{stetson}. A total of 1,500 artificial stars was added
to the $F110W$- and $F160W$-band frames in several
steps of 150 stars each. These had magnitudes and colors in the range
$18 \leqslant F110W \leqslant 25$ and $0.5 \leqslant (F110W-F160W) \leqslant 1.5$.
Stars were considered as recovered if they were found in both $F110W$ and
$F160W$ with magnitudes not exceeding $0\fm75$ brighter than the initial,
injected ones. The outcome of the completeness tests is shown in
Fig.~\ref{f:cmpl_nic2}.

\subsection{WFPC2}

\begin{figure}[htb]
\vbox{\includegraphics{dio.fig6a.eps}}
\vspace{2.9cm}
\vbox{\includegraphics{dio.fig6b.eps}}
\vspace{2.9cm}
\vbox{\includegraphics{dio.fig6c.eps}}
\vspace{4.7cm}
\caption{{\tt ALLSTAR} residuals of the
fitting of the PSF models
{\it versus} magnitude for WFPC2. Only stars included in the final
photometric list of NGC~4214 are plotted.
\label{f:ST2}
}
\end{figure}

\begin{figure}[tbh]
\vbox{\includegraphics{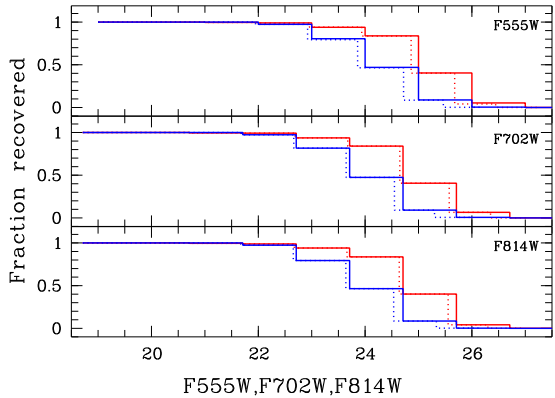}}
\vspace{5.7cm}
\caption{Completeness levels of the WFPC2 photometry based on
artificial
star tests with colors $[-0\fm5\leqslant(F555W-F814W)\leqslant1\fm5]$.
The solid and dashed lines are the completeness histograms as a function of
the input and detected magnitude
for the star-forming complexes and the ``field" areas
(see Fig.~\ref{f:blue_red_star}).
\label{f:cmpl_wfpc2}
}
\end{figure}

The WFPC2 observations of program 6569\footnotemark[\value{footnote}]
were obtained on 1997 July 22.
The $F555W$ ($V$, total exposure time 1200~s), $F702W$ ($R$, 1000~s), and
$F814W$ ($I$, 1200~s), $F656N$ ($H\alpha$; 1500~s), and
$F502N$ ([$O\,${\sc iii}], 1600~s)
frames were centered on (J2000) R.A. $12^{\rm h}15^{\rm m}39\fs95$
and Dec. $+36\degr19\arcmin35\farcs8$.
WFPC2 consists of one
$35\arcsec \times 35\arcsec$ PC chip and three $77\arcsec \times 77\arcsec$
chips, WF2,~3,~4. The center of NGC~4214-{\sc i}
was placed on the WF3 chip, and
NGC~4214-{\sc ii} was located on the WF2 chip. There is
spatial overlap between the WF2 and NIC2 pointings, which we capitalize on
for a multi-wavelength study of NGC~4214-{\sc ii}. 

The raw frames were processed with the standard WFPC2 pipeline.
The processed frames were then separated into images
for each individual CCD, and trimmed of the vignetted
regions using the boundaries recommended in the WFPC2 Handbook.
The locations of bright stars were then measured on each set (one for each
combination of filter, position and CCD) to evaluate
any systematic changes in position during the sequence
of observations. None were found. To check for possible saturation 
of the brightest supergiants we also reduced and
performed photometry on short exposure time (100 s) frames in the $F555W$
and $F814W$ bands. Comparison with the long exposure frames showed a good
agreement.
Unfortunately, the WFPC2 observations were performed without dithering, so we
were unable to enhance the stellar PSF, using the drizzling procedure.
We combined the images after cleaning them for bad pixels and
cosmic\--ray events.

The $F555W$, $F702W$, $F814W$ images were processed with
{\tt DAOPHOT}/{\tt ALLSTAR}. To avoid contamination by nebular emission, 
we performed $F555W$-band photometry on
$F502N$-subtracted images, and we performed  $F702W$-band photometry on
$F656N$-subtracted images.
Transformation to STMAG was made using header keywords (WFPC2 photometry
cookbook).
We used \citet{Holtzman95} to transform instrumental
magnitudes into the standard $V$, $R_{c}$, and $I_{c}$ system.
Figure~\ref{f:ST2} shows the error distributions for the
photometry.

We performed completeness tests with the goal of assessing the
accuracy of the TRGB distance determination. We used only those areas on
the WFPC2 chips which are outside of the star-forming regions, 
and the completeness tests are consequently only relevant to
those areas. For the WFPC2 completeness tests, we added a total of
5,000 artificial stars to the $F555W$, $F702W$ and $F814W$ frames of
NGC~4214. We added artificial stars with magnitudes and colors in the range
$20 \leqslant F814W \leqslant 27$,
$-0\fm5 \leqslant (F555W-F814W) \leqslant 1\fm5$ or
$-0\fm3 \leqslant (F702W-F814W) \leqslant 0\fm5$,
in several steps of 500 stars each. 
The completeness curves for red
stars are shown in Fig.\ref{f:cmpl_wfpc2}.

\section{Results} \label{results}

\subsection{Extinction and foreground contamination} \label{foreground}

The Galactic coordinates of NGC~4214 are $l_{II}~=~160\fdg25,
b_{II}~=~78\fdg07$. In the maps of
\citet{burstein}, the extinction is $A_B~=~0\fm000$, but in the IRAS/DIRBE map of
\citet{Schlegel98} it is $A_B = 0\fm094$. 
We adopt the latter value. 
With $R_V~=~3.1$ and the extinction law of \citet{cardelli},
Galactic foreground extinction is: $A_V~=~0\fm073$, $A_R~=~0\fm058$,
$A_I~=~0\fm042$, $A_J~=~0\fm020$, and $A_H~=~0\fm013$.

Internal extinction is very uncertain for the central part of NGC~4214.
There are large discrepancies in the values derived with
different spectroscopic methods \citep[e.g.][]{Maiz98,Mas99}.
Using the continuum slope of the ultraviolet spectrum and assuming that
Galactic foreground extinction is negligible, \citet{Leitherer96}
found a color excess of $E(B-V)=0\fm09$ for NGC~4214-{\sc i}. After
correction for Galactic extinction from \citet{Schlegel98} this value
corresponds to $E(B-V)=0\fm07$.
But optical spectroscopic measurements of Balmer emission lines give
an internal reddening for NGC~4214-{\sc i}/{\sc ii} from
$E(B-V)=0\fm09$ \citep{Sargent91} up to $E(B-V)=0\fm28$
\citep{Mas99}.

According to these data, dust appears concentrated near the boundaries of the
ionized NGC~4214-{\sc i} region, affecting mainly nebular emission
lines, while the stellar continuum itself is located in a region relatively
free of dust and gas \citep{MacKenty00}.
The less affection by extinction of the stellar component than the ionized
gas seems to be common for some of WR/Irregular galaxies \citep{Kunth86}

The WFPC2 images clearly show several dust lanes, which are more
pronounced across the bright NGC~4214-{\sc i} complex. Unfortunately,
WFPC2 observations in the $U$-band were
not deep enough to derive stellar photometry we could use to 
estimate the extinction toward field stars.  Therefore, we have not attempted
to correct these data for internal extinction.  

Thanks to its high galactic latitude, we expect the number of foreground 
stars toward NGC~4214
to be very small. This has been verified using \citet{Ratnatunga85},
who determined 
the Galactic stellar density toward globular clusters. The nearest cluster
to NGC~4214 is Pal~4 ($l = 202\degr$, $b = 72\degr$). The number of stars
expected down to a limiting magnitude $V=26$ is 7 stars~arcmin$^{-2}$.
WFPC2 covers $\sim 5$ arcmin$^2$, so the model predicts $\sim 35$ Galactic
stars. For the NIC2 image we predict less
than one star for the entire field down to the limiting magnitude.
We found a few background galaxies in the WFPC2 frames, and 
rejected them from the star list.

\subsection{Colour-magnitude diagrams} \label{CM}
\subsubsection{NICMOS}

\begin{figure*}[htb]
\vbox{\includegraphics{dio.fig8a.eps}
       \includegraphics{dio.fig8b.eps}
\vspace*{6.8cm}
}
\caption{Colour--magnitude diagrams for
$\sim 1,500$ stars
in the HST Vegamag system, corrected for foreground extinction. 
The dashed lines show the position of TRGB.
\label{f:CM_nic2}
}
\end{figure*}

A CMD of the near-IR photometry is displayed as Fig.~\ref{f:CM_nic2},
in terms of magnitudes in the VegaMAG system.
There are $\sim 1,500$
stars detected in both the $F110W$ and $F160W$ filters
with {\tt ALLSTAR} parameters in
the intervals $-2 \leqslant SHARP \leqslant 2$  and $CHI \leqslant 5$.
The CMDs are characterized by a strong red plume and a weak blue plume.
Most of the stars in the red plume appear at faint magnitudes;
this is the red tangle which contains the RGB.
Above this tangle is a population of
possible asymptotic giant branch (AGB) and RSG stars.
Note the pronounced fall off in the number of stars redder than
$F110W-F160W \ga 1.2$, an almost vertical ``border''.
We saw a similar feature in our data and modeling for Mrk~178 \citep{rsl00}.

\subsubsection{WFPC2}

\begin{figure*}[htb]
\vbox{\includegraphics{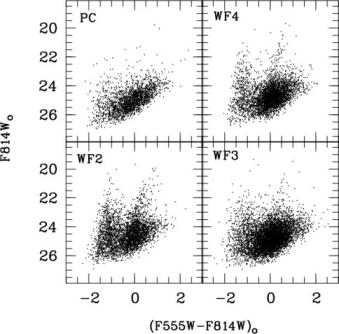}
\vspace*{13.0cm}
}
\caption{Extinction-corrected CMDs for the different chips
of WFPC2 using the photometry in $F555W$ and $F814W$. Spatial variations
in the stellar content are immediately apparent from the varying strengths
of the blue and red plumes.
\label{f:CMFWPC2}
}
\end{figure*}

\begin{figure*}[htb]
\vbox{\includegraphics{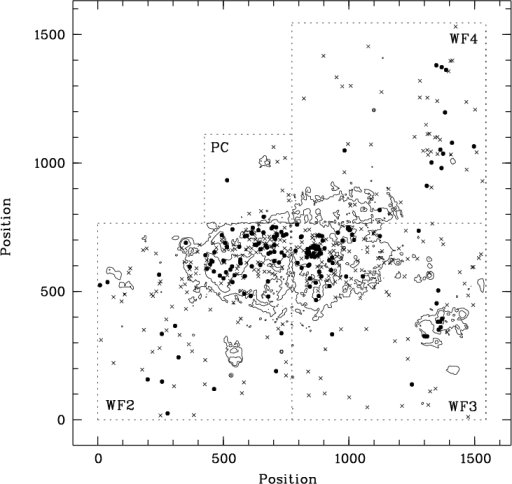}
\vspace*{12.0cm}
}
\caption[blue_red_star.ps]{Distribution of the brightest blue
(filled circles) and red stars
(crosses) on the WFPC2 image of NGC~4214. Contours of H$\alpha$
flux are overlaid. We used these isophotes to
separate the regions with active star formation from older stellar
populations, and refer to the latter are the ``field" areas.
\label{f:blue_red_star}
}
\end{figure*}

The optical CMDs which result from single-star photometry on the WFPC2 chips are
shown in Fig.~\ref{f:CMFWPC2}, in terms of magnitudes in the STMAG system.
In determining the distance and metallicity based on the TRGB method,
we rely on the assumption that as we go away from the star-forming regions
of NGC~4214, star counts become less contaminated by young and intermediate-age
stars. Such stars can overlap the RGB above its tip and tend to smear
out the TRGB ``edge'' in the luminosity function (LF;
Sakai \& Madore \citeyear{Sakai99}).
Figure~\ref{f:blue_red_star} illustrates how we chose areas outside
of the star-forming regions based on H$\alpha$-isophotes.
These areas also exhibit less crowding and less internal
extinction than the regions of active star formation. 
The CMDs of these ``field'' regions
corrected for Galactic extinction
are shown in Fig.~\ref{f:CM_wfpc2}.
There are $\sim 20,000$ stars detected in both the $F555W$ and $F814W$ filters
in all four WFPC2 frames
with $SHARP$ and $CHI$ {\tt ALLSTAR} parameters in
the intervals $-0.5 \leqslant SHARP \leqslant 0.5$  and $CHI \leqslant 1.5$.
About 23,000 point sources are detected after merging the $F702W$ and $F814W$
star lists, and applying the same selections.

\begin{figure*}[hbt]
\vbox{\includegraphics{dio.fig11a.eps}
      \includegraphics{dio.fig11b.eps}
\vspace*{6.8cm}
}
\caption[CMAll_vi.ps,CMAll_ri.ps]{The $(V-I)_0$--$I_0$ and
$(R-I)_0$--$I_0$ CMDs
for $\sim 15,000$ and $\sim 23,000$ stars from the ``field" parts of NGC~4214
used in the TRGB distance determination.
\label{f:CM_wfpc2}
}
\end{figure*}

\subsubsection{Optical and near-IR data combined}

\begin{figure*}[tb]
\vbox{\includegraphics{dio.fig12a.eps}
      \includegraphics{dio.fig12b.eps}
      \includegraphics{dio.fig12c.eps}}
\vspace*{5.4cm}
\caption{Optical--near-IR color-magnitude and
color-color diagrams
of NGC~4214-{\sc ii} showing stars which have been found in $V$, $I$, $J$, and
$H$. An extinction vector of $A_V=1$ is shown in the Color-color diagram.
\label{f:CC}
}
\end{figure*}

The NIC2 images are situated well within the WF2 images of NGC~4214.
We cross-identified sources found in both cameras by transforming
the WF2 coordinates into the NIC2 system. 
We then merged our photometry lists requiring a positional source coincidence
of better than $0.1\arcsec$, and investigated the distribution of stars on a
variety of two-color diagrams and CMDs (Fig.~\ref{f:CC}).
There are $\sim 600$ stars in the [$(I-H)_0$,$I_0$] CMD, and $\sim 300$ stars
in the [$(V-H)_0$,$V_0$] CMD.

Few simultaneous optical--near-IR stellar CMDs of external galaxies are
available to date. Comparing our data with those of VII~Zw~403 (SHGC99) reveals 
a great deal of similarity. Note that the red plume continues to
dominate in the [$(I-H)_0$,$I_0$] CMD, but 
the blue and red plumes are about equally well populated 
in the [$(V-H)_0$,$V_0$] CMD. The
[$(V-H)_0$,$V_0$] CMD has a color baseline of 8~mag, and shows
a pronounced separation of the blue and red plumes.

We also show the distribution of stars on the  
$(V-H)_0$, $(V-I)_0$ two-color diagram. 
The effect of differential reddening could in
principle be detected on a two-color diagram, but the distribution
of stars in these particular colors corresponds closely with
the direction of the reddening vector.  Recall that the observations
in $U$ did not provide useful photometry.   

\section{Discussion}

\subsection{ Distance and metallicity} \label{distance}

\begin{table}
\begin{center}
\caption{{\sc The Tip of the Red Giant Branch}\label{t:TRGB}}
\vspace{0.3cm}
\begin{tabular}{lcc}
\tableline
\tableline
Band      & $m_{\rm TRGB}$ & $M_{\rm TRGB}$\\
\tableline
$F814W_{0,STMAG}$ \dotfill   & $24\fm31$ & $-2\fm82$\\
$F814W_{0,VegaMAG}$ \dotfill & $23\fm05$ & $-4\fm08$\\
$I_0$ \dotfill     & $23\fm10$ & $-4\fm03$\\
$F110W_0$ \dotfill & $22\fm85$ & $-4\fm28$\\
$J_0$ \dotfill     & $22\fm55$ & $-4\fm58$\\
$F160W_0$ \dotfill & $21\fm86$ & $-5\fm43$\\
$H_0$ \dotfill     & $21\fm77$ & $-5\fm36$\\
\tableline
\end{tabular}
\end{center}
\end{table}

\begin{table}
\begin{center}
\caption{{\sc Errors for TRGB Distance}\label{t:errTRGB}}
\vspace{0.3cm}
\begin{tabular}{ll}
\tableline
\tableline
Source  & Error\\
\tableline
{\bf NICMOS}&\\
Residuals from PSF fitting \dotfill & $\pm0.05$\\
Tip measurement in VII~Zw~403\dotfill & $\pm0.10$\\
Tip measurement in NGC~4214\dotfill & $\pm0.10$\\
Galactic extinction \dotfill & $\pm0.02$\\
Internal extinction \dotfill & $\pm0.05$\\
{\bf Total random errors \dotfill} & {\bf $\pm0.16$}\\
&\\
RR Lyrae distance \dotfill & $\pm0.15$\\
Absolute magnitude of TRGB \dotfill & $\pm0.10$\\
Transformation to $F110W$ and $F160W$ \dotfill & $\pm0.05$\\
Photometric zero points\dotfill & $\pm0.05$\\
{\bf Total systematic errors \dotfill} & {\bf $\pm0.19$}\\
&\\
{\bf WFPC2}&\\
Residuals from PSF fitting \dotfill & $\pm0.10$\\
Tip measurement \dotfill & $\pm0.10$\\
Galactic extinction \dotfill & $\pm0.03$\\
Internal extinction \dotfill & $\pm0.05$\\
{\bf Total random errors \dotfill} & {\bf $\pm0.15$}\\
&\\
RR Lyrae distance scale \dotfill & $\pm0.15$\\
Absolute magnitude of TRGB \dotfill & $\pm0.10$\\
Transformation to $I$ \dotfill & $\pm0.03$\\
Photometric zero points\dotfill & $\pm0.02$\\
{\bf Total systematic errors \dotfill} & {\bf $\pm0.18$}\\
\tableline
\end{tabular}
\end{center}
\end{table}

The absolute magnitude of the TRGB gives
an estimate of distance with precision and accuracy similar to
that of the Cepheid method \citep{lee,Bellazzini01}.
It is slightly dependent on metallicity. A metallicity determination
is made to correct for this dependence. The $I$-band TRGB method has been
successfully applied to a wide range of dwarf galaxies, using both
ground-based and HST photometry. We first
present a distance to NGC~4214 based on this well established technique.
It is also desirable, in some cases, to be able to extend the TRGB
method to the near-IR. In SCHG99, we combined $V$, $I$, $J$,
$H$ photometry obtained with HST WFPC2 and NIC2 in order to
bootstrap a calibration for the near-IR TRGB from the optical TRGB
for the case of the metal-poor Blue Compact Dwarf galaxy VII~Zw~403.
The location of the TRGB in $J$ and $H$ depends more strongly on
metallicity than does the location of the TRGB in $I$. Furthermore, $J-H$
color is not a good indicator of metallicity. The second part
of this section involves deriving a distance to NGC~4214 from the
calibrations which we derived for VII~Zw~403. This assumes that the
RGB of NGC~4214 has a low metallicity similar to that of VII~Zw~403.
Indeed, the distance so derived agrees with that
determined from the optical TRGB.
Finally, we also derive an independent calibration for the 
NGC~4214 $H$-band TRGB using as input the $I$-band TRGB and the
metallicity derived from $V-I$ color. We discuss the dependence
of the $H$-band TRGB on metallicity in the following subsection. 

The location of the TRGB is commonly found by computing a
luminosity function (LF) along the red plume, and then applying
an edge-detecting Sobel filter \citep{sakai}.
Fig.~\ref{f:SLF} shows the LFs and Sobel-filtered
LFs in $F110W$, $F160W$, and $I$.
We computed the LFs by counting stars lying inside a bin of
$\pm 0\fm12$. The central
value was varied in steps of $0\fm04$ to reduce
the dependence of our results on the particular choice of bin center.

\begin{figure}[hbtp]
\vbox{\includegraphics{dio.fig13a.eps}}
\vspace{5.1cm}
\vbox{\includegraphics{dio.fig13b.eps}}
\vspace{5.1cm}
\vbox{\includegraphics{dio.fig13c.eps}}
\vspace{5.4cm}
\caption{Smoothed luminosity
function (solid lines) and edge-detection
filter output (dotted lines) for $F110W$, $F160W$ and $I$ bands.
Vertical lines indicate the position of TRGB.
\label{f:SLF}
}
\end{figure}

For the near-IR LFs,  we selected all stars in the color
interval $0\fm75 \leqslant (F110W-F160W)_0 \leqslant 1\fm4$.
In the optical, our selection proceeded
as follows. The photometric limits in the [$(R-I)_0$,$I_0$] CMD are deeper than in
the [$(V-I)_0$,$I_0$] CMD, while the color resolution is better in $(V-I)_0$.
Therefore, while we used the [$(V-I)_0$,$I_0$] CMD for deriving the
metallicity, we selected stars on the [$(R-I)_0$,$I_0$] CMD for deriving the
position of the TRGB. To obtain the magnitude of the TRGB we used
the LF of the stars in the color interval
$0\fm2<(R-I)_0\leqslant 1\fm2$. Note that only the ``field'' regions
which are free of ionized gas emission are used for the optical TRGB determination.
Finally, an edge-detecting Sobel
filter $[-2,0,+2]$ was applied to the LFs;
the position of the TRGB is identified with the highest peak in
the respective filter output function.
The TRGB magnitudes so derived are listed in Table~\ref{t:TRGB}.

The color of the TRGB is necessary to calculate the bolometric
correction to be used in the $I_{\rm TRGB}$-distance calibration.
It is taken as the median color index of the stars with
$22\fm7 \leqslant I \leqslant 22\fm9$.
We find $(V-I)_{\rm TRGB}=1\fm42$ which,
when corrected for external extinction, yields $(V-I)_{\rm TRGB,0}=1\fm39$.
For the RGB stars the average stellar metallicity can be obtained from the
index $(V-I)_{\rm-3.5}$, which is the color index of the RGB half a magnitude
below the TRGB \citep[see][]{dacosta,lee,Bellazzini01}.
We believe that the TRGB is at  $I_{\rm TRGB,0}=23\fm1$ (see below).
The median color between $I_0=23\fm5$ and $I_0=23\fm7$, which we use to
determine
$(V-I)_{\rm -3.5,0}$, is $1\fm34\pm0\fm28$.
Using the calibration by \citet{lee}, we obtain a metallicity
[Fe/H]$\simeq-1.7$ dex ($Z_{\sun}/50$), and an internal abundance spread
with a total range of -1.0 dex from the intrinsic color
width of the RGB. This width could result
from metallicity or age variations in an evolving RGB population, 
but other factors such as internal extinction might also contribute.

Using the calibration by \citet{lee}, we obtain $M_{I_0,{\rm TRGB}}=-4\fm03$.
We then derive a distance modulus $(m-M)_0=27\fm13$, corresponding
to $2.7$ Mpc. The error budget is given in Table~\ref{t:errTRGB}.
Note that we estimate the error in foreground extinction adopting a
16\% accuracy of the \citet{Schlegel98} maps.  For the error in internal
extinction, for which we have no direct measurement, 
we estimate 0.1 mag. 
Our total error is $\pm0\fm15\pm0\fm18$ (random \& systematic) or
$\pm0.3$~Mpc.

The near-IR distance modulus of NGC~4214 may be derived using the
calibration which was established for VII~Zw~403
in \citet[SHGC99]{rsl99},

\centerline{$M_{F110W_0, {\rm TRGB}}=-4\fm28\pm 0\fm10 \pm 0\fm18$}
\noindent and\\
\centerline{$M_{F160W_0, {\rm TRGB}} =-5\fm43 \pm 0\fm10 \pm 0\fm18$}

\noindent where the first error is the statistical error and is dominated by
how well we can determine the location of the TRGB in VII~Zw~403, and the
second one is the systematic error primarily due to the RR~Lyrae distance
calibration of the TRGB \citep[see][SCH98]{rsl98}.

Using this calibration, we determine the following distance moduli for
NGC~4214

\begin{center}
$(m-M)_{F110W_0} = 22\fm85 + 4\fm28 = 27\fm13\pm 0\fm16 \pm 0\fm19$
\end{center}
\noindent and
\begin{center}
$(m-M)_{F160W_0}= 21\fm86 + 5\fm43 = 27\fm29\pm 0\fm16 \pm 0\fm19$.
\end{center}

\noindent
The error budget is again listed in Table~\ref{t:errTRGB}.

The RGB stars of NGC~4214 have a mean
metallicity that is close to that of VII~Zw~403, [Fe/H]=-1.7 $vs.$
-1.9, respectively (both are based on $V-I$ color). Therefore, 
we expect the near-IR 
TRGB to be close to the optical TRGB. This is indeed observed.


We also note that a purely theoretical calibration 
of the TRGB method was presented by
\citet{Cassisi97}, who used evolutionary models of stars.
If we adopt this calibration instead of that of \citet{lee},
we obtain $M_{I_0,{\rm TRGB}}=-4\fm27$, and thus a distance modulus of
$(m-M)_0=27\fm37$. This corresponds to a distance of 3.0~Mpc,
about 10\% farther.


As discussed above, 
distance estimates to NGC~4214 have been a matter of some debate. 
A comprehensive discussion
of the distance to NGC~4214 was given in \citet{lsaz97},
who adopt a value of 4.1 Mpc
based on ground-based photometry of the brightest blue stars, but
they noted that this value is an upper limit.
The average magnitude of the three brightest stars in a galaxy is
a simple, frequently used method to estimate the distance. However, the
brightest supergiants method has large errors.
For illustrative purposes, we derive the distance to NGC~4214 by this method
also. The HST data resolve the supergiant populations somewhat
better than the ground-based data do, and hence the apparent magnitudes of
the brightest stars should decrease (because of less blending), as should the
distance modulus. The mean apparent
magnitude of the brightest blue stars can be estimated from our $VI$ photometry
(using now the CMD of all 4 WFPC2 chips, not just that of the ``field"
regions). In order to make a direct comparison with \citet{lsaz97}, we used
the transformation $(B-V)=0.83\cdot(V-I)$ obtained from blue [$(V-I)\leqslant0\fm6$]
standard stars of \citet{Landolt}. Applying this to the stars with
$(V-I)\leqslant0\fm6$,the mean apparent magnitude of the three brightest
BSGs is $\langle B(3B) \rangle\simeq18\fm78$.
Using the standard relation
$(m-M)_0(B)=1.51\cdot \langle B(3B)\rangle -0.51\cdot B_{\rm T}-A_B+4.14$
from \citet{ikar_ntik94}, an extinction $A_B=0\fm09$, and the total apparent blue magnitude
from the RC3, $B_{\rm T}=10\fm24$, we get
$(m-M)_0=27\fm18$. This is close to the TRGB distance.
However, using the magnitude of three brightest red supergiants (RSG),
$\langle V(3R) \rangle_0 \simeq20\fm49$
and the calibration of \citet{ikar_ntik94},
$\langle M_V(3R) \rangle =-7\fm7$, we derive $(m-M)_0=28.19$,
which is about 1~mag larger than the distance modulus via the TRGB method or
more than 50\% larger in distance.

\subsection{Calibrating the $H$-band TRGB}\label{calibr_H-TRGB}

\begin{figure}[htb]
\vbox{\includegraphics{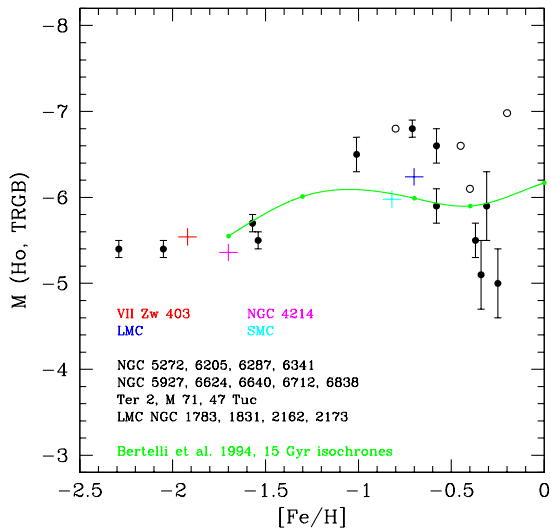}
\vspace{7.0cm}
}
\caption{
Absolute H-band magnitude at the TRGB versus metallicity. The crosses
mark the positions of the galaxies discussed in this paper.
\label{f:H_TRGB}
}
\end{figure}

In this section, we use the new data on NGC~4214 to provide an
additional  
calibration 
of the absolute $H$-band magnitude at the TRGB as a distance indicator.
This study began in SHGC99, and continued in \citet{hopp01}.
Fig.~\ref{f:H_TRGB} represents an update, with the data point for NGC~4214
included.

Knowing the location of the optical TRGB of NGC~4214,
we can predict where we should have found the $H$-band TRGB.
The transformed $H$ magnitude at the TRGB using the transformations from
SHGC99 is
\begin{center}
m$_{H_0, {\rm TRGB}}$~=~21\fm8~$\pm$~0\fm1. \\
\end{center}

\noindent
Given that the distance modulus based on the well-established $I$-band
TRGB method is $27\fm13\pm0\fm23$, we predict the absolute $H$ magnitude at the
TRGB to be

\begin{center}
M$_{H_0, {\rm TRGB}} = -5\fm3 \pm 0\fm3$ \\
\end{center}

\noindent
This is the new calibration included in Fig.~\ref{f:H_TRGB}. 
 
The TRGB in $H$ offers the advantage of being about
2 mag brighter, on average, than the TRGB in $I$. Unfortunately,
theoretical stellar models indicate a complicated relationship between 
[Fe/H] and the TRGB in $H$. The $H$-band magnitudes from the Padova
isochrones (Bertelli et al. 1994, see also Fagotto et al. 1994)
for 15 Gyr old stars for the largest available metallicity range are shown
in Fig.~\ref{f:H_TRGB}. We also compiled $J$, $H$ data of globular clusters
in the Milky Way and the LMC from the literature (see SHGC99).
This yields an observational
measure of the dependence of absolute $H$-band TRGB magnitude
on metallicity. The major difficulty of this approach is that frequently, the
empirical RGBs of clusters are not sufficiently populated near
the tip to provide a reliable tip magnitude.

\begin{figure}[tbp]
\vbox{\includegraphics{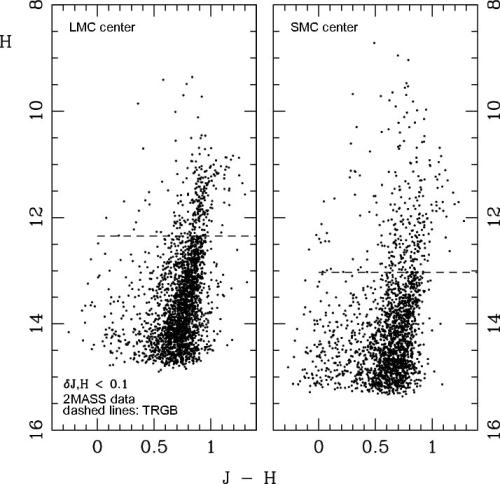}
\vspace{7.0cm}
}
\caption{
Near-IR CMDs for the LMC and the SMC based on 2MASS data of their central
regions
\label{f:LSMC_CMD}
}
\end{figure}

The release of the Two Micron All Sky Survey (2MASS) data made
it possible to include near-IR data for field stars in the LMC and the SMC in our
comparison \citep[see also][]{hopp01}. The data for the LMC and the SMC were
taken from the 
point source catalog of the second incremental data
release\footnote{http://www.ipac.caltech.edu/2mass/releases/second}.
Figure~\ref{f:LSMC_CMD} shows
CMDs for the central fields ($600\arcsec$ radius) of the LMC and the SMC
in the 2MASS system, which is similar to the CIT/CTIO system.
The major advantage of this
data set is that there are such large numbers of RGB stars detected that
we need not worry about stochastic fluctuations in the luminosity functions
near the TRGB due to small number statistics, but only about AGB contamination. 
Our best estimates for the TRGB magnitudes are $H_{\rm TRGB}=12\fm35$ for
the LMC and $H_{\rm TRGB}=13\fm03$ for the SMC, with a read-off error of
about $\pm0\fm05$.

The foreground extinction
according to \citet{Schlegel98} are A$_H=0\fm043$ for the LMC, and
A$_H=0\fm021$ for the SMC, so that the extinction-corrected
TRGB magnitudes are 
$H_{\rm 0, TRGB} \approx 12\fm3$ for the LMC, and
$H_{\rm 0, TRGB} \approx 13\fm0$ for the SMC. In order to place these data
points onto Fig.~\ref{f:H_TRGB}, we need distances and metallicities for
the LMC and SMC.

The SMC and LMC distances have been a matter of some debate in the recent literature. 
Distances and metallicities of the LMC and SMC using single-star near-IR photometry 
were recently derived by \citet{Cioni00}. We decided to adopt their
values here. They find m-M = $18\fm55$ ($0\fm04$ formal, $0\fm08$ systematic)
for the LMC, and m-M = $18\fm99$ ($0\fm03$ formal, $0\fm08$ systematic) for
the SMC. This results in
M$_{H_{\rm 0, TRGB}}\approx -6\fm24$ for the LMC and
M$_{H_{\rm 0, TRGB}}\approx -5\fm98$ for the SMC.
We assume [M/H] = -0.7 for the LMC, and [M/H]= -0.82 for the SMC
from \citet{Cioni00} to represent the Fe/H values of the RGB stars in
Fig.~\ref{f:H_TRGB}.

A comparison of empirical RGB-based distance and metallicity determinations
of VII~Zw~403, NGC~4214, the LMC and the SMC, with the 15~Gyr isochrones
in the Padova library shows that there is excellent agreement between
observation and theory. Fig.~\ref{f:H_TRGB} supports our conclusion from
SHGC99 and \citet{hopp01}
that $\langle$M$_{H_{\rm 0, TRGB}} \rangle = -5\fm5 \pm 0\fm1$ is indeed
a good approximation to use for deriving
the distances of galaxies with very low metallicities.

\subsection{Distance-dependent global parameters}\label{parameters}

The distance dependent global parameters of NGC~4214 are summarized in
Table~\ref{t:dist_par}.
At our newly determined distance of 2.7 Mpc, 
an angular scale of $1\arcsec$ corresponds to a physical scale of 13~pc.
The diameter of NGC~4214 (D$_{25}$ from the RC3) is $\sim8\farcm5$. 
This corresponds to a linear size of 6.7 kpc. The total apparent blue
magnitude from the RC3, $B_{\rm T}^0 = 10\fm14$,
translates into a total absolute magnitude of $-16\fm99$,
or a blue luminosity of about $8.3\cdot10^8$~L$_{\sun}$
(for M$_{B,\sun}$=5.41). 

At our new distance, the H{\sc i} flux
measured by \citet{Huchtmeier85} corresponds to an H{\sc i}
mass of $5.5\cdot10^{8}$ ${\cal M}_{\sun}$. Note that the ratio of
the H{\sc i} mass to the
blue luminosity (in solar units) is 0.7; this value is typical
compared with the compilation of dwarf-galaxy ${\cal M}_{\rm HI}$/L$_B$
ratios given in \citet{Huchtmeier97}.
A total molecular mass for NGC~4214 estimated from $^{12}CO(1-0)$ flux
by \citet{Walter01} for the corrected distance gives
$2.2\cdot10^6 {\cal M}_{\sun}$.

The star formation rate (SFR) is one of the most important parameters for 
studying the evolution of galaxies.
When individual young stars are unresolved the star formation properties
of galaxies come from integrated light measurements in the ultraviolet
(UV), far-infrared (FIR), 1.4~GHz-radio or nebular recombination lines
\citep[e.g.][]{Condon92,Kennicutt98}. These diagnostic methods are based on
population synthesis modeling.
\citet{MacKenty00} measured the H$\alpha$
fluxes of the prominent H{\sc ii}-regions. We derive a total luminosity, 
corrected for foreground extinction only, of 
L(H$\alpha$)$\simeq 1.2 \cdot 10^{40}$ ergs s$^{-1}$. A similar value is
derived from \citet{Martin98}.
Following \citet{Hunter86}, who adopt the Salpeter IMF from 0.1
to 100 ${\cal M}_{\sun}$, we find the SFR $0.08$ ${\cal M}_{\sun}$ yr$^{-1}$.
Considering the unknown internal extinction and any emission outside
the WFPC field of view, 
we regard this as a lower
limit to the SFR. Using the calibration from \citet{Kennicutt98} instead, we
derive a SFR of $0.09 {\cal M}_{\sun}$ yr$^{-1}$.
These values are consistent with that derived by \citet{Gallagher84},
$\sim 0.1 {\cal M}_{\sun}$ yr$^{-1}$ (after rescaling to our distance).

\citet{Fanelli97} used the total FUV 1500\AA\phantom{1} flux to estimate
the SFR of NGC~4214. They used starburst models for two limiting cases:
a 3 Myr burst with IMF slope = -1.35 and a mass range 
$0.1{\rm (extrapolated)} < {\cal M}_{\sun} <100$, $Z = 0.4 Z_{\sun}$;
and a continuous star formation model
with the same parameters otherwise. Only correction for Galactic extinction
was taken into account. We rescaled this SFR to the new distance and derive
a value of $\sim 0.08 {\cal M}_{\sun}$~yr$^{-1}$, in good agreement with the
value from the H$\alpha$ flux.
Using the 1.4~GHz flux of the central body from
\citet{Allsopp79}, $S_{1.4}=50\pm10$~mJy,
and the calibration from \citet{Haarsma00}, we
derive a SFR of $\sim 0.06 {\cal M}_{\sun}$~yr$^{-1}$. The IRAS $60 \mu$m flux of
17.87~Jy, with the calibration of \citet{Cram98}, yields a SFR of
$\sim 0.17 {\cal M}_{\sun}$~yr$^{-1}$.

\begin{table}[hbt]
\begin{center}
\caption{{\sc Distance-dependent global parameters}\label{t:dist_par}}
\vspace{0.3cm}
\begin{tabular}{lrr}
\tableline
\tableline
Parameter                           &   Value   &               Ref. \\
\tableline
Distance modulus                    & $27\fm13$ &                  1 \\
Distance                            & 2.7 Mpc   &                  1 \\
Linear size (D$_{25}\simeq8\farcm5$)& 6.7 kpc   &                  2 \\
M$_{B,{\rm total}}$ ($B_{\rm T}^0=10\fm14$) & $-16\fm99$ &         2 \\
L$_{B}$ (for M$_{B,\sun}=5\fm41$)   & $8.3\cdot10^8$~L$_{\sun}$ &  2 \\
${\cal M}_{\rm HI}$        & $5.5\cdot10^{8} {\cal M}_{\sun}$ &    3 \\
${\cal M}_{\rm CO}$        & $2.2\cdot10^6 {\cal M}_{\sun}$ &       4 \\
L (FUV)                    & $6.3 \cdot 10^{26}$ ergs s$^{-1}$ Hz$^{-1}$& 5 \\
L (H$\alpha$)              & $1.2 \cdot 10^{40}$ ergs s$^{-1}$ & 6,7 \\
L (1.4 GHz)                & $4.4 \cdot 10^{26}$ ergs s$^{-1}$ Hz$^{-1}$& 8 \\
L ($60 \mu$m)              & $1.4 \cdot 10^{42}$ ergs s$^{-1}$ &   9 \\
SFR (H$\alpha$)            & $0.09 {\cal M}_{\sun}$ yr$^{-1}$  &  10 \\
SFR (UV)                   & $0.08 {\cal M}_{\sun}$ yr$^{-1}$  &  11 \\
SFR (Radio)                & $0.06 {\cal M}_{\sun}$ yr$^{-1}$  &  12 \\
SFR (FIR)                  & $0.17 {\cal M}_{\sun}$ yr$^{-1}$  &  13 \\
\tableline
\end{tabular}
\tablerefs{(1) This paper; (2) RC3; \\
       (3) \citet{Huchtmeier85}; (4) \citet{Walter01}; \\
       (5) \citet{Fanelli97}; (6) \citet{MacKenty00};\\
       (7) \citet{Martin98}; (8) \citet{Allsopp79};\\
       (9) \citet{IRAS}; (10) \citet{Kennicutt98};\\
       (11) \citet{Fanelli97}; (12) \citet{Haarsma00};\\
       (13) \citet{Cram98}}
\end{center}
\end{table}

How reliable are these SFRs? Apart from discrepancies that might
arise from the applicability of the calibrating relationships from
luminosity to SFR \citep[e.g.][]{Cram98}, the main factor causing differences
in the derived SFRs
of NGC~4214 is that fluxes in different wavelengths refer to different 
measurement beams or apertures, which in turn covered different galaxy areas.
Numerous star formation complexes in NGC~4214 are traced out to large
galactocentric distances. The radio and the IRAS data have the largest beam sizes and
included a larger fraction of the galaxy body (and possibly some of external background
sources too) than did the UV and the H$\alpha$ measurements. Therefore, the
fact that the total SFRs agree to within a factor of three suggests that the two
main complexes, NGC~4214-{\sc i} and {\sc ii}, dominate the young stellar population
of NGC~4214. 

Evidence for large cavities in the interstellar medium of NGC~4214 around
sites of massive star formation has been found from ionized and neutral
hydrogen observations \citep{Maiz99,MacKenty00,Walter01}. The cavities
appear to have been created by stellar winds from young massive stars and
supernova explosions. As was mentioned by \citet{Leitherer96},
this may cause an underestimate of the current SFR.

Another possible reason for differences in the derived SFRs is that each
of the four indicators we use is associated with a different time scale.
Therefore, a limitation, which is shared by all of the methods, is
the dependence of the derived SFRs on the assumed star formation history
(SFH).
Most of these calibrations use a model of continuous SFR over a time scale
of $10^8$ years or longer, and are not really acceptable for young starburst
galaxies \citep{Kennicutt98}. For example,
burst models for a 9-million-year-old population yield SFRs that are
57\% higher than those with the assumption of constant SFR
\citep[e.g.][]{Leitherer95}. 

\citet{Gallagher84} explored the star formation history of NGC~4214 
using three parameters that sample its SFR at different epochs: (1) the mass
of a galaxy in the form of stars, which measures the SFR integrated over a
galaxy's lifetime; (2) the blue luminosity, which is dominated primarily by
stars formed over the past few billion years; and (3) the Lyman continuum
photon flux derived from H$\alpha$ luminosity, which gives the current
($\la 10^8$~yr) SFR. They found that NGC~4214 is among the burst/postburst
irregular galaxies and assumed a scenario of roughly constant star
formation over the recent few hundred million years. In paper~II, we shall
present our own models of the SFH of NGC~4214 based on the photometry
discussed in this paper. The data have good time resolution and will
provide a detailed SFH for the past Gyr. In addition, we can
also (weakly) constrain the birthrate parameter based on the modeled data. 
Note that the mere appearance
of the CMDs of NGC~4214 is already very supportive of a quasi-continuous mode
of star formation, as inferred by \citet{Gallagher84}.

Note that
NGC~4214 contains sufficient neutral gas to maintain a rate of
$0.1 {\cal M}_{\sun}$~yr$^{-1}$ for a Hubble time.


\subsection{Stellar Content and its Spatial Variation}\label{absCMD}

\begin{figure}[tbh]
\vbox{\includegraphics{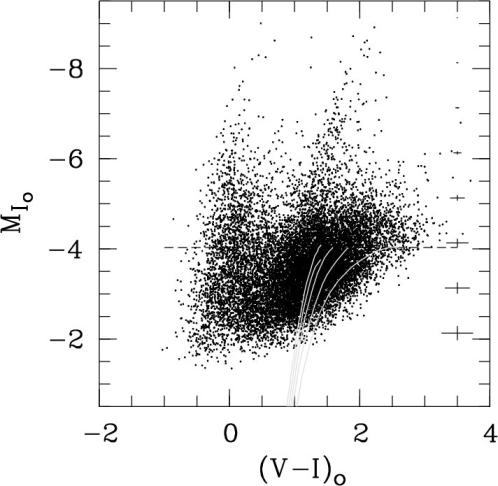}
\vspace{6.8cm}
}
\caption{[$(V-I)_0, M_{I_0}$] CMD of the field-star
populations of NGC~4214 (the same as that shown in Fig.~\ref{f:CM_wfpc2}).
Overplotted are giant branches of the standard globular clusters
M~15 ([Fe/H]$=-2.17$), NGC~6397 ([Fe/H]$=-1.91$), M~2 ([Fe/H]$=-1.58$),
NGC~1851 ([Fe/H]$=-1.29$), and 47~Tuc ([Fe/H]$=-0.71$), from
\citet{dacosta}.
\label{f:CMD_Miv}
}
\end{figure}

\begin{figure}[tbh]
\vbox{\includegraphics{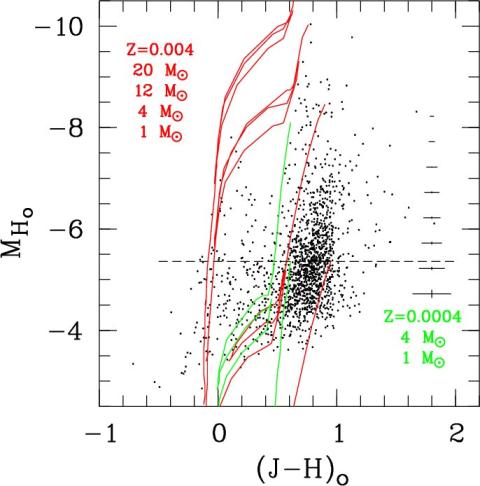}
\vspace{6.8cm}
}
\caption{$M_{H_0}$ {\it versus} $(J-H)_0$ CM diagram of NGC~4214-{\sc ii}
complex. Stellar tracks for metallicities of Z=0.004 and Z=0.0004 from the Padova
library are overplotted. 
\label{f:CMD_JH}
}
\end{figure}

Figure~\ref{f:blue_red_star} illustrates the spatial positions
of the brightest blue and red stars.
We selected the brightest stars as having
$M_{I_0} \leqslant -6\fm0$
and $(V-I)_0 \leqslant 0\fm5$ for the blue stars (for a total
of 163 stars)  and
$(V-I)_0 \geqslant 1\fm0$ for the red stars (for a total of 313). 
The stellar populations in NGC~4214 show the
younger stars to be spatially clumped.
Overall, most of the brightest
blue and red stars are concentrated in the two main star formation
complexes.
The largest stellar density occurs near the center of the H$\alpha$ cavity,
Super Star Cluster (SSC)
I-A in \citet{MacKenty00} notation. Photometry in this region has a low
completeness, due to high crowding of the stellar images.
NGC~4214 has a substantial young population also outside of its major
star-forming complexes. Concentrations of bright stars are seen outside
the central area, possibly tracing a continuation of the blue 
bar/arm-like structures seen in the Isaac
Newton Image (Fig.~\ref{f:int_ima}).

To examine the stellar content and its spatial variations across the HST field
of view, we use the H$\alpha$ flux outer isophotes
(see Figure~\ref{f:blue_red_star}) to separate the stellar populations
into ``field'' regions which are free from ionized gas emission, and
the more complicated regions inside the star-forming complexes.

Fig.~\ref{f:CMD_Miv} shows the [($V-I$)$_0$,$M_{I_0}$] CMD of
these ``field'' regions in terms of absolute magnitudes. 
This CMD has characteristics typical of
the CMDs of other dwarf galaxies, with evidence for both old and
young populations. The main feature of this CMD
is the concentration of stars that extends between $0\fm7\la(V-I)_0\la2\fm0$
and $M_{I_0}\ga-4\fm5$. This corresponds to the ``red tangle"
which is anticipated to be populated by RGB, AGB and blue loop (BL)
old and intermediate-age stars.
There is also a considerable population of blue stars ($(V-I)_0\la 0.7$)
in this CMD, suggesting that there are young stars in the field. 
Note that the object identified with the
most luminous BSG is actually part of the field-star population, 
outside the H$\alpha$ regions.  This is also true for the
most luminous RSGs. The dominant feature of the CMD
is the red tangle, with a stellar density which is clearly higher
than that in the blue plume and in the top of the red plume.
In Fig.~\ref{f:CMD_Miv}
we superimpose on the CMD the globular cluster ridgelines of \citet{dacosta}.
This serves to further illustrate the large width of this feature.
In principle, the red tangle could contain stars with ages all the way
up to a Hubble time. The dominance of the red tangle in the field of
NGC~4214 suggests that the galaxy possesses an underlying sheet of
intermediate-age and possibly ancient stars.

\begin{figure}[tbh]
\vbox{\includegraphics{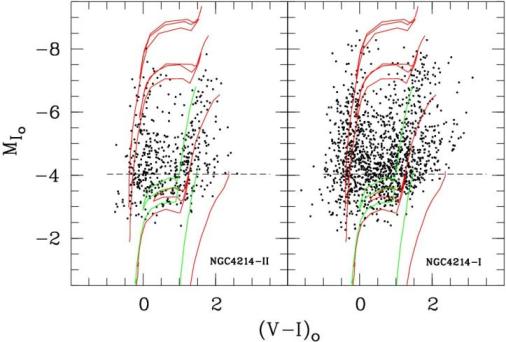}
\vspace{5.0cm}
}
\caption{CMDs of stars lying in the star formation
regions NGC~4214-{\sc i} and -{\sc ii}. Overplotted are the tracks for same stellar
masses and metallicities as those
used in Fig.~\ref{f:CMD_JH}.
\label{f:CMvi-I-II}
}
\end{figure}

The NIC2 observations encompass NGC~4214-{\sc ii}.
Figure~\ref{f:CMD_JH} shows the near-IR CMD transformed into $J$ and $H$.
Despite the fairly large uncertainty in such transformations (see SHG99) an
advantage of transforming the data into the $JHK$ system is that our CMDs
and LFs can be more readily compared with the few available ground-based
data of similar galaxies in the near-IR (see \S~\ref{calibr_H-TRGB}). We also
overplot a selection of stellar evolutionary tracks.
We find that a range of stellar phases is included in the CMD of this
region. The $12 {\cal M}_{\sun}$ track ($Z=0.004$) provides an envelope to the
blue and red supergiant populations detected in this region. However, the
CMD also clearly shows the underlying older stars of NGC~4214, which we
represent with $1 {\cal M}_{\sun}$ tracks at $Z=0.004$ and $Z=0.0004$.
In addition, there is a substantial population of AGB and thermally
pulsing, or TP-AGB, stars. We give intermediate-mass
tracks for $4 {\cal M}_{\sun}$, noting that they terminate at the first
thermal pulse. Sources with $(J-H)_0 \approx 1\fm5$ are probable carbon-rich
TP-AGB stars (carbon stars).

The [$(V-I)_0$,$M_{I_0}$] CMDs
of the unsaturated stars lying in the complexes NGC~4214-{\sc i}
($\sim1,200$ stars) and NGC~4214-{\sc ii} ($\sim500$ stars) are shown in
Fig.~\ref{f:CMvi-I-II}.
A comparison of the CMDs inside the star-forming regions with
the CMD of the field-star population indicates substantial similarities
between the two. The most obvious difference is the almost complete
absence of red tangle stars with $V-I$ colors between about 1 and 2 mags.
in the CMDs of the complexes. This corresponds to a lack of stars between
the low and high metallicity tracks for a $1 {\cal M}_{\sun}$ star.
We attribute
this paucity of red tangle stars to severe crowding. Experience has
shown that the red background sheet is not seen in the presence
of high crowding inside star-forming regions \citep[e.g.][]{rsl98}.
The CMD of NGC~4214-{\sc i} is similar to that
of NGC~4214-{\sc ii}. Both show bright blue supergiants near $M_I\sim-8\fm0$. 
The $12 {\cal M}_{\sun}$ track ($Z=0.004$) is a good envelope to the
distribution of the brightest stars on the optical CMD of NGC~4214-{\sc ii}.
This agrees with our finding from the near-IR. The CMD of NGC~4214-{\sc i}
exhibits a number of stars which are more luminous, and are located
near the $20 {\cal M}_{\sun}$ track ($Z=0.004$). However, since the total
number of stars on the CMD of NGC~4214-{\sc i} is higher than that
on the CMD of NGC~4214-{\sc ii}, the probability of finding rare,
luminous stars is also higher here. We would not want to derive
an age difference between the two regions based on the CMDs in
Fig.~\ref{f:CMvi-I-II} alone;  a careful analysis of the crowding properties of
the two regions would be necessary. 
In NGC~4214-{\sc i}
there are a few extremely red stars with $(V-I)_0 \geqslant 2\fm5$, which are
TP-AGB stars. Such extremely red stars appear to be missing in
NGC~4214-{\sc ii}. Again, since the TP-AGB phase is a short-lived
phase of stellar evolution, a smaller number of stars in NGC~4214-{\sc ii}
is expected based on the smaller total number of stars on its CMD.


The work of \citet{MacKenty00} allows us to give H$\alpha$-derived SFRs normalized to
area for the two prominent SF complexes, NGC~4214-{\sc i} and {\sc ii}.
For NGC~4214-{\sc i} we derive $2.1 \cdot 10^{-7} {\cal M}_{\sun}$ yr$^{-1}$ pc$^{-2}$,
and for NGC~4214-{\sc ii} we determine $3.4 \cdot 10^{-7} {\cal M}_{\sun}$ yr$^{-1}$ pc$^{-2}$,
using the calibration of \citet{Kennicutt98}. These two complexes also
resolve at 1.4~GHZ in the FIRST, Faint Images of the Radio Sky at Twenty
centimeters \citet{Becker95}, database. For NGC~4214-{\sc i} we find
$5.4 \cdot 10^{-7} {\cal M}_{\sun}$ yr$^{-1}$ pc$^{-2}$,
and for NGC~4214-{\sc ii} we calculate
$11 \cdot 10^{-7} {\cal M}_{\sun}$ yr$^{-1}$ pc$^{-2}$.
The agreement between the two estimators is quite good. Both identify NGC~4214-{\sc ii}
as the complex with the higher SFR/area.

To summarize, we find that stars populate the CMD of NGC~4214 throughout all of the
major phases of stellar evolution above our detection limits. 
The relative numbers of stars in different phases
can in principle give the variation of the SFR with time and spatial location. However,
this requires a detailed modeling of the CMDs, which is not included in
this paper. 
Since all of the major stellar phases are represented on the CMDs,
we conclude that the star formation was more or less continuous 
\citep[gasping scheme, see][]{Tosi01} in recent times (last 1-2 Gyrs).
 Given that young stars are found in
the field, we note that the star-formation activity
is not exclusive to the high-surface-brightness, central complexes. 
The SFH can be traced
back to at least 1-2~Gyr ago, the minimum
age compatible in the Padova library with a distinct TRGB feature. Stars
could have
formed even earlier in the history of NGC~4214, but we cannot
tell from the CMD. We can also use the spatial distribution of
stars as an age tracer. RGB and AGB stars are seen throughout the
entire visible disk of NGC~4124. 
While massive, young 
stars are also scattered throughout the disk of NGC~4214, 
there is
a distinct concentration of luminous stars in the core, from which
most of the H$\alpha$ luminosity originates as well.
Therefore, most of the on-going activity takes place near the
center of the distribution of the older stars.

\subsection{Environment}\label{environment}

\begin{figure*}[tbp]
\vbox{\includegraphics{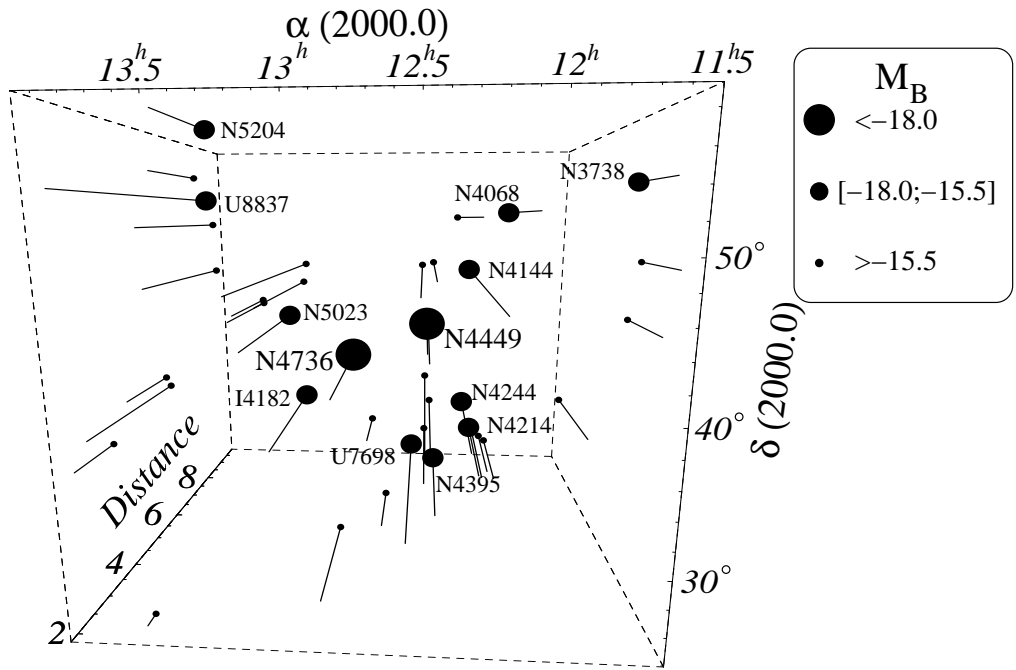}
\vspace*{8.7cm}
}
\caption{A scaled 3-D representation
[($\alpha$; $\delta$; {\it Distance to LG\/}]
of the Canes Venatici cloud. The size of circle corresponds to the
galaxy's absolute magnitude.
Lines are proportional to the galaxy's distance. Names are given
for the bright dominant galaxies using the notation 
'N' for 'NGC', 'I' for 'IC', and 'U' for 'UGC'.
\label{f:CVn_map}
}
\end{figure*}

Our new distance estimate places NGC~4214 on the near side (or possibly in the
foreground) of the CVn Cloud of galaxies.
One of more prominent galaxies of the CVn Cloud, NGC~4214 has some
possible nearby companions. The dwarf galaxy DDO~113, and the
starforming irregular galaxy NGC~4190, have similar radial velocities as
NGC~4214 and angular separations from NGC~4214 of $\la 30\arcmin$.
\citet{Allsopp79} suggests that since the H{\sc i}
emissions of NGC~4214 and NGC~4190 extended towards each other, they might
be experiencing a physical interaction.

The spatial structure of the CVn Cloud is poorly known. 
The distribution of galaxies within this complex is quite scattered,  
and it is not clear whether it is gravitationally
bound \citep{ikar_dio98}.
Fig.~\ref{f:CVn_map} illustrates the three-dimensional distribution of potential
CVn member galaxies on the sky.  We obtained the catalog
from Karachentsev (1999, private communication), and determined
distances to the LG barycenter using photometric distances in the literature
\citep{ikar_ntik94,Georgiev97,lsaz97,lsaz98,ntik_ikar98,ikar_dio98,rsl99,mcrone00,rsl00,rsl01}.
The standard deviation in radial velocities of  
CVn cloud galaxies is about +65 km/s. The
photometric distance estimates have a standard deviation of
$\sim1.7$~Mpc. 
More accurate distance determinations, similar to that which we
presented in this paper for NGC~4214, are needed for the other possible member
galaxies before we can provide more stringent constraints on the
spatial structure and kinematics of the CVn cloud.

\section{ Concluding Remarks} \label{conclusions}

We present $V$, $R$, $I$, $J$, $H$ photometry of resolved stars
in the star-forming dwarf galaxy NGC~4214.
The optical and near-IR data show 
stellar populations which are distinct in structure and color:
a low-surface-brightness stellar population of intermediate-age/old stars which
is co-spatial with a sparse young field-star population, and
an inner ($\sim 1$~kpc) high-surface-brightness young population
which is co-spatial with the most prominent H-{\sc ii} regions.

NGC~4214 has a complex physical structure.
Whereas the spatial distribution of red stars
is almost uniform and traces the large stellar disk, the distribution of
brightest blue stars is clumpy and follows the central bar, along with 
arm-like chains of stellar associations and H-{\sc ii} regions.

The CMDs are populated by stars in a range of evolutionary phases:
young MS, BSG, and RSG stars, young and intermediate-age BL and AGB stars,
TP-AGB stars (amongst them candidates for carbon stars), and intermediate-age 
and possibly old RGB, stars. These populations trace star formation 
back as much as 1-2~Gyr ago. In paper~II, we shall present solutions for the
SFH of NGC~4214 based on model CMDs. The appearance of the CMDs of NGC~4214
is suggestive of the ``gasping", rather than the ``bursting", mode of star
formation.

Assuming that the red tangle outside of the most prominent regions of
active star formation is dominated by old RGB stars, we derive a stellar
metallicity
of about $Z_{\sun}/50$ ([Fe/H]~$\sim -1.7$ dex).
The ionized gas metallicity (O abundance) of about $Z_{\sun}/4$ suggests
some metallicity evolution.
However, we note that the RGB of NGC~4214 has a substantial
width, which could be due to a spread in metallicity or age, plus internal
extinction.

We derive a distance of $2.7$ Mpc from the TRGB.
Using the optical and near-IR
CMDs we define more precisely the relationship between
the metallicity of the RGB stars and their tip in the near-IR.
For this we use our new results for NGC~4214, determinations of SHGC99 for
VII~Zw~403, and 2MASS data obtained for the Magellanic clouds.
A comparison of empirical data with the 15~Gyr isochrones in the Padova
library indicates that there is an excellent agreement between observation
and theory.

Our distance estimate leads to a revision of all
distance-dependent global parameters (Table~\ref{t:dist_par}).
As our value is
about 1.5 times closer than was adopted in many previous papers, 
absolute luminosity
estimates should be reduced by a factor of $\sim 2$. 
Given that its revised absolute blue magnitude is $-16\fm99$, NGC~4214
definitely qualifies as a dwarf galaxy. 
We find an H$\alpha$-based SFR of
$\sim 5 \cdot 10^{-4} {\cal M}_{\sun}{\rm yr}^{-1}$
\citep{MacKenty00,Kennicutt98} for NGC~4214-I-As, the complex used for
the spectral comparison with Lyman-break galaxies \citep{Steidel96b}.
In conclusion, NGC~4214 exhibits global
parameters that are rather typical for a dwarf Irregular galaxy.
Its star formation properties are much less extreme than previously
thought \citep[e.g.][]{Thronson88,Fanelli97}, a fact that must be taken into
account when it is compared with high-redshift galaxies.


\begin{acknowledgements}
Work on this project was supported through HST grants to RSL and MMC
(project 7859). UH acknowledges financial support from SFB~375.
This publication makes use of the Royal Greenwich
Observatory/Isaac Newton Group Archives,
the NASA/IPAC extragalactic database (NED)
which is operated by the Jet Propulsion Laboratory, Caltech, under contract
with the National Aeronautics and Space Administration,
and data products from the Two Micron All Sky Survey,
which is a joint project of the University of Massachusetts and the Infrared
Processing and Analysis Center/California Institute of Technology, funded
by the National Aeronautics and Space Administration and the National Science
Foundation.
We are grateful to Dr.~P.B.~Stetson for his comments to NICMOS stellar
photometry, Dr.~A.~Hopkins for his help estimating the SFRs and
Dr.~L.N.~Ma\-ka\-ro\-va for providing her paper.

\end{acknowledgements}


{}


\end{document}